# Examining Different Placement Strategies for Indoor Environmental Quality Sensors in Office Environments

**Riccardo Talami** [a], **Xinhao Hu** [ab], **Ilyas Dawoodjee** [a], **Ali Ghahramani** [a] *

a. Department of the Built Environment, National University of Singapore, Singapore, 117566.
b. College of Civil Engineering, Hunan University, Changsha, 410081, China.

**ABSTRACT**

Collecting Indoor Environmental Quality (IEQ) data from an occupant's immediate surroundings can provide personalized insights for healthy environmental conditions aligned with occupant preferences, but effective sensor placement for data accuracy and reliability has not been thoroughly explored. This paper explores various positioning of IEQ multi-sensing devices at individual workstations in typical office settings, aiming to identify sensor placements that most accurately reflect the environmental conditions experienced by occupants. We examined five unique positions close to an occupant (above and below the monitor, right side of the desk, ceiling, and chair backrest), two orientations, and three desk locations characterized by different lighting levels, thermal and airflow conditions. Data on temperature, humidity, carbon dioxide ($CO_2$), particulate matters ($PM_1$, $PM_{2.5}$, $PM_{10}$), illuminance, and sound were collected over a 2-week longitudinal experiment, followed by short-term experiments simulating common pollution events such as coughing and sneezing. Principal Component Analysis, Spearman's rank correlation, $R^2$, and Mean Absolute Error were applied to identify the position and orientation that best captures the most information and matches breathing zone measurements. It was found that above the monitor position, facing the occupant, best captures the IEQ conditions experienced by the occupant.

## 1. Introduction

The deployment of sensors to capture what building occupants are experiencing has been limited, as it is challenging to obtain real-time access to Indoor Environmental Quality (IEQ) parameters in close proximity to occupants. However, failing to capture poor environmental conditions, such as inadequate lighting levels, poor air quality, or uncomfortable temperatures (ASHRAE, 2017), can lead to health problems such as headaches, eye strain, and respiratory issues (Hedge and Dorsey, 2013; Wargocki and Wyon, 2017). This can also affect the occupants' productivity and cognitive function (Wyon, 2004; Fisk, 2000).

With the advancement of sensor technologies, accuracy, and reduced costs, real-time and continuous data can be achieved with a wide variety of sensors for IEQ monitoring (Ródenas García et al., 2022; Saini et al., 2021), and a large amount of data can be leveraged to investigate thermal, visual, and acoustic comfort, and indoor air quality (IAQ) in office buildings. Capitalizing on the personalized data collected from occupants in the immediate surroundings of their everyday environments (e.g., office workstation) can offer a deeper understanding of individuals' comfort needs, as they represent what occupants are actually experiencing. As such, personalized information on real-time and long-term monitoring data can support intelligent and condition-specific indoor environmental control and management (Parkinson et al., 2019a), leading to the creation of occupant-centric environments.

The placement of sensors is crucial for effective IEQ monitoring, as their incorrect positioning can compromise their performance (Suryanarayana et al., 2021), resulting in unreliable data collection that may ultimately cause occupants' discomfort. Currently, the deployment of sensors for IEQ monitoring often relies on industry best practices or ease of installation (Yun & Licina, 2023). Moreover, guidelines and standards such as ASHRAE Standard 55 (ASHRAE, 2017) and 62.1 (ASHRAE, 2022), WELL v2 (V.2 Well, 2018), and RESET v2 (RESET® standard, 2022) provide recommendations for sensor positioning in buildings to ensure effective monitoring of IEQ. However, these standards often focus on room-level or space-level sensor placement rather than capturing the specific conditions that affect individual comfort and health. Similarly, Cheng et al. (2020) and Yun et al. (2018) explicitly point out gaps in sensor positioning research around occupants in office environments, particularly noting the need for more granular studies in such settings.

Furthermore, both standards and previous studies consider only individual IEQ parameters, relying on sensors dedicated to measuring a single IEQ parameter. However, sensing packages that integrate multiple IEQ parameters into a single device are increasingly emerging for continuous IEQ monitoring applications (Fritz et al., 2022; Parkinson et al., 2019a, 2019b; Tiele et al., 2018). This is due to the fact that integrating multiple sensors within a single device can reduce installation costs, sensor expenses, and power use. When deploying multi-sensor devices, conflicting positioning needs can arise due to varying sensor requirements, leading to compromises between the most suitable placement for one sensor and avoiding inaccuracy in others (Z. Chen et al., 2018). Therefore, it is essential to ensure that positioning of IEQ sensing packages strikes balance between multiple metrics and ensuring accurate data collection. With the paradigm shift towards occupant-centric buildings which places the well-being, comfort and productivity of building occupants at the forefront of design and operation, building guidelines are expected to progressively incorporate factors related to occupants, leading to a wider adoption of sensors to gauge the comfort level and satisfaction that individuals experience in the indoor environment (O'Brien et al., 2020).

Table 1 breaks down 13 studies from the past 3 years that explore indoor environmental monitoring.

Table 1. Studies on indoor environmental monitoring and sensor placement in the past 3 years.

| Monitored environment | Sensor type | Environmental parameter | Research method | Sensor placement | Analysis method | Duration | Reference |
|---|---|---|---|---|---|---|---|
| Room | A P-DustMonit and a VOCs sensor | PM and VOC | Experimental measurement | Not specified | Data readings | Single day | Fermo et al. 2021 (Fermo et al., 2021) |
| House setting | Stationary environmental sensors, personal exposure and IoT low-cost sensor | $CO_2$, temperature, relative humidity, size-resolved particle number concentration, VOC and SVOC | Experiment measurement | Furniture of each room | Regression analysis | 3 days | González Serrano et al. 2023 (González Serrano et al., 2023) |
| Office room | Handheld aerosol monitor | PM | Experiment measurement and numerical simulation | Table in the middle of the room | Data readings | 6 days | Fu et al. 2022 (Fu et al., 2022) |
| Climate chamber | Low-cost air quality monitors and single sensors | PM, $CO_2$, VOC, temperature, and relative humidity | Experiment measurement | Table in the middle of the room | Standard deviation, root mean square error | 1 hour | Zheng et al. 2022 (Zheng et al., 2022) |
| Kitchen | Low-cost air quality monitors | $CO_2$, CO, $NO_2$, $O_3$, PM2.5 | Experiment measurement | On top of kitchen cabinets | Data readings | 7 days | Tryner et al. 2021 (Tryner et al., 2021) |
| Test room, chamber and bedroom | All-in-one low-cost sensors | $CO_2$, PM, VOCs, CO, temperature, relative humidity and light | Experiment measurement and field study | Table in the chamber and 1 meter above the ground | Linear regression models | 2 weeks | Fritz et al. 2022 (Fritz et al., 2022) |
| Controlled chamber | Seven stationary air quality sensors | $CO_2$ and PM | Experiment measurement | 2 on the HVAC exhaust, 2 walls, 2 desk | Significance, correlation, multiple linear regression and error analysis | 60 mins x 4 dynamic conditions, 70- and 40-mins x 2 static conditions | Yun et al. 2023 (Yun & Licina, 2023) |
| Building | Stationary sensor | Temperature | Data driven | Each wall | Statistical test and greedy algorithm | Not applicable | Suryanarayana et al. 2021 (Suryanarayana et al., 2021) |
| Office room | Multi-sensors | Temperature, relative humidity, $CO_2$, passive infrared and light | Experiment measurement | Walls, ceiling, desk | Data fusion and correlation analysis | 2 weeks | Azizi et al. 2021 (Azizi et al., 2021) |
| Office floor | Simulated sensors | Temperature and $CO_2$ | BIM and CFD simulation | Ceiling | Genetic Algorithm | Not applicable | Cheng et al. 2022 (Cheng et al., 2022) |
| Climate chamber | Low-cost Arduino-compatible sensors | Temperature, relative humidity and $CO_2$, | Experiment measurement | Table in the room | Correlation | Not reported | Pereira et al. 2022 (Pereira & Ramos, 2022) |
| Seminar room | Simulated sensors | $CO_2$ | CFD modelling and simulation | Ceiling | Airflow and $CO_2$ spatial distribution | Not applicable | Mou et al. 2022 (Mou et al., 2022) |
| Educational building | Simulated sensors | Temperature | CFD simulation | Walls and ceiling | Mean and standard deviation analysis | Not applicable | Chen et al. 2021 (C. Chen & Gorlé, 2022) |

Low-cost sensors are increasingly being employed in studies monitoring a wide array of environmental parameters (Ali et al., 2016; Fritz et al., 2022; González Serrano et al., 2023; Karami et al., 2018; Pereira & Ramos, 2022; Tryner et al., 2021; Zhang et al., 2021; Zheng et al., 2022). However, the focus of these studies primarily revolves around the development of sensors and their evaluation against high-precision reference sensors in terms of readings calibration and their monitoring capacities. Therefore, all the aforementioned studies overlooked the impact of sensor positioning on the monitored data. In fact, they solely relied on empirical methods for sensor placement, devoid of any analysis to derive the recommended sensor location. Azizi et al. (2021) collected occupancy, $CO_2$, temperature, humidity, and illuminance data from 3 single-occupant offices for a qualitative description of the effects of sensor positioning on environmental monitoring, highlighting the importance of sensor placement for data reliability.

While some studies have examined the impact of sensor location on environmental monitoring and have recommended sensor locations, they primarily focused on IAQ metrics (Cheng et al., 2022; Mou et al., 2022; Suryanarayana et al., 2021; Yun and Licina, 2023, Rackes et al., 2017). As such, the most suitable sensor placement for monitoring comprehensive IEQ (thermal, lighting, IAQ, acoustic) is unexplored. Furthermore, sensor positioning is typically investigated at the room level, whether through experiments or simulations. However, there is a need to move beyond this approach and customize sensor placement for personalized monitoring. This tailored approach ensures accurate representation of environmental conditions experienced by individuals. Finally, there is a lack of studies that provide generalizable findings validated across different scenarios.

Suryanarayana et al. (2021) proposed a data-driven methodology to identify placement of sensors in a multi-zone building. While this method has the advantage of being data-driven yet computationally efficient, it requires detailed building models and large datasets to evaluate the number and position of sensors. Mou et al. (2022) modelled and simulated airflow and $CO_2$ spatial distribution of a seminar room to derive the placement of $CO_2$ sensors for data reliability. Although they found that positions near the ceiling allowed for more effective air mixing, thus accurately capturing areas of elevated $CO_2$ concentration, the adoption of complex and computationally intensive Computational Fluid Dynamics (CFD) numerical simulations prevents the adoption of this methodology for practical applications. Cheng et al. (2022) developed a simulation-based methodology to optimize temperature and $CO_2$ sensor positioning in a multi-zone office environment using a genetic algorithm (GA) and machine learning. The generated sensor placements are in line with the requirements from LEEDv4 and can be adapted to the number of occupants in the rooms. However, the complexity of the models and expertise required to generate the solutions can affect usability in practice. Yun and Licina (2023) investigated sensor placement with respect to $CO_2$, $PM_{2.5}$, and $PM_{10}$ for personal exposure detection. Employing a Decision Tree Classifier, they identified that positions that best represented $CO_2$ exposure in both static-sitting and dynamic-standing conditions were the area behind the reference occupant's sidewall and the reference occupant's desk, while the most suitable stationary PM sensor locations were either the sidewall behind the reference occupant or the ceiling-mounted ventilation exhaust. However, the identified sensor positions vary across scenarios and are case-specific, hence hindering the generalizability of the findings. In addition, recommended sensor positioning concerning all three IAQ metrics is unclear. Finally, the methodology involves analyzing sensor readings from different locations compared to reference concentrations in the breathing zone. While this is a valuable approach for IAQ sensing device, it is hardly applicable to other types of environmental parameters such as temperature, humidity, sound, and illuminance levels, as these metrics lack a corresponding reference for comparison (breathing zone). In summary, there is a lack of generalizable recommendations on sensor package placement and positioning in office buildings for personalized IEQ monitoring of thermal, visual, acoustic, and IAQ parameters.

## 1.1. Research gaps, Objectives and Contribution to knowledge

Despite advancements in IEQ sensor technology, significant research gaps remain regarding sensor placement in office environments. Existing studies have primarily focused on room-level conditions, neglecting the localized environmental conditions experienced by occupants at their workstations. In addition, research has yet to fully explore the validity of multi-sensor devices capable of simultaneously measuring thermal comfort, IAQ, lighting levels, and acoustic comfort. Standards such as ASHRAE Standard 55 (ASHRAE, 2017) and 62.1 (ASHRAE, 2022), WELL v2 (V.2 Well, 2018), and RESET v2 (RESET® standard, 2022) also fail to adequately address the growing trend of integrating multiple sensors into a single device, which presents conflicting placement requirements.

This paper addresses these gaps by:

- Evaluating the effectiveness of different sensor positionings in capturing real-time IEQ conditions at individual workstations.

- Proposing placements of multi-sensor devices that account for multiple IEQ parameters.

Specifically, the study evaluates five sensor placements (above and below the monitor, the right side of the desk, ceiling-mounted, and on the backrest of the chair), two orientations (horizontal orientation where the sensors face the ceiling, and vertical orientation where the sensors directly face the occupant) across three desk locations characterized by varying environmental conditions, collecting data on temperature, humidity, $CO_2$, $PM_1$, $PM_{2.5}$, $PM_{10}$, illuminance, and sound levels. Through a combination of continuous monitoring and controlled experiments simulating occupants' coughing and sneezing events, this research aims to identify the sensor placement that most accurately reflects conditions experienced by occupants.

The findings of this study will contribute to a deeper understanding of the relationship between sensor positioning and occupant comfort, providing actionable insights for intelligent building management systems. This research challenges previous research, which focus on room-level placement and single-parameter monitoring, and advocates for more personalized sensor placement strategies. In doing so, the results of this work could inform future updates to IEQ guidelines, particularly for integrated multi-sensor devices deployed in proximity to occupants.

This paper is structured as follows. Section 2 presents the methodological workflow adopted in this study, including the description of the experimental setting, sensor package design and validation, the investigated sensor positions, the experimental procedure, and the data analytics. Section 3 presents the findings, followed by the discussions in Section 4, and limitations and suggestions for future research in section 5. Finally, section 6 concludes the paper with a breakdown of the results.

## 2. Methodology

This section details the methodological workflow employed in this study (Figure 1).

Section 2.1 presents the experimental setting. The experiments were conducted in a living laboratory equipped with three desks and designed to simulate a typical office environment. This setup allowed for variations in lighting, thermal conditions, and airflow patterns, which were key factors in determining sensor placements that most accurately reflect the environmental conditions experienced by occupants.

Section 2.2 details the development and validation of the IEQ sensor package. A sensor package was developed to monitor monitor IAQ (Carbon dioxide, Particulate matter, and Total volatile organic compound), lighting (Illuminance), thermal (Temperature and Humidity), and acoustic (Sound) parameters. A two-fold analysis was carried out to assess the accuracy of the sensor package and validate its compliance with standards.

Section 2.3 presents the sensor placement for personalized monitoring. Fifteen stationary IEQ sensing packages were placed across the three desks in the test environment to generalize the findings. Five monitoring locations were selected at varying heights and distances from the occupant: above and below the computer monitor, on the desk to the right side of the table, attached to the ceiling, and on the backrest of the chair. These locations were chosen to capture diverse environmental conditions from the occupant's perspective. Measurements were collected for two sensor orientations: (1) horizontal, with the sensor facing the ceiling, and (2) vertical, with the sensor facing the occupant.

Section 2.4 describes the longitudinal study and controlled experiments. A two-week longitudinal study was conducted with three human subjects to assess real-time environmental variations in the office setting. In addition, two controlled experiments were carried out to simulate air pollutant concentration increases. A particle dispersion machine was used to mimic human coughs and sneezes.

Section 2.5 outlines the data analytics. Principal Component Analysis (PCA) was employed to analyze the data collected from the five monitoring positions, three desks, and two sensor orientations. The PCA identified the most suitable sensor position by determining which location captured the most variance and relevant information concerning the occupant. Furthermore, Spearman's rank correlation analysis ($\rho$), goodness of fit ($R^2$), and Mean Absolute Error (MAE) were used to compare the sensor data with ground truth measurements from the particle dispersion machine's nozzle, or "breathing zone." These analyses helped to identify the sensor position that most accurately matched the ground truth data during the short-term experiments.

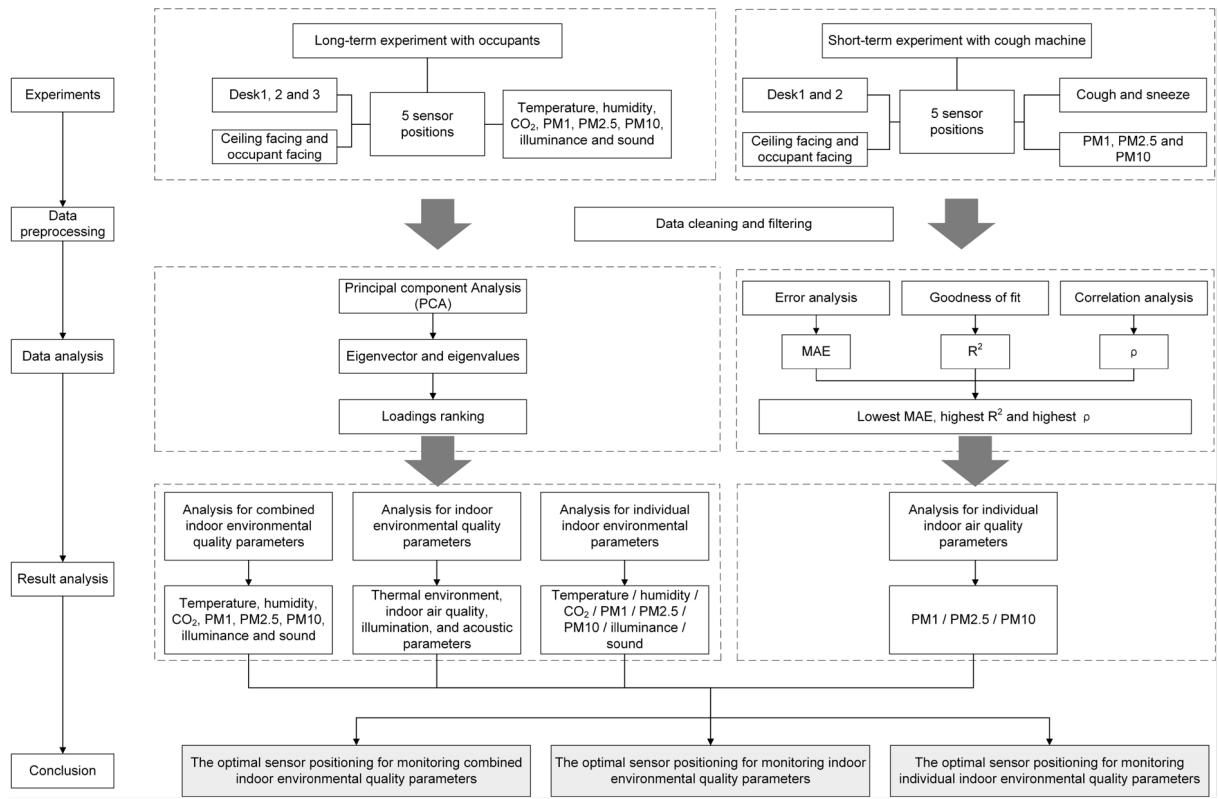

Figure 1. Methodological workflow for the identification of the recommended sensor positioning.

## 2.1 Experimental setting

The experiments were conducted in a controlled climate chamber turned open-plan living laboratory with a floor area of 75 m$^2$ and a volume of 195 m$^3$ (Figure 5), situated on the third floor of a building. The HVAC system controlled the room air temperature and relative humidity at 26 ± 1°C and 60 ± 5%, respectively, measured across 5 ceiling sensors in the room. A mixing strategy involving 100% of outdoor air ensured acceptable ventilation rates, supplied and exhausted through 6 swirl type diffusers in the ceiling. Tile ceiling lights with neon tubes provide a constant level of illuminance throughout the day supplied by a 1.5m tall ribbon window spanning the entire length of the office. The open-plan office is equipped with desks, screens and chairs and is used daily by researchers during working hours (10am-6pm), ensuring dynamic occupancy conditions. For the experiments, 3 desks were used to investigate the impact of their spatial positioning within the open-plan multi-occupancy office on the IEQ metrics. In fact, the layout, design, and configuration of the monitored office can significantly impact the distribution of environmental parameters. Factors such as the presence of HVAC diffusers and exhausts, windows, and occupant activities can create spatial variability in IEQ metrics. Hence, occupants sitting at different desk locations experience variations in lighting, thermal conditions, and airflow patterns. The desks selected for the experiments represent common scenarios occurring in office settings. As depicted in Figure 5, desk 1 is located near the office door and is placed under the ceiling diffuser; hence the lighting levels are mainly dictated by the artificial lights, outdoor heat gains are minimum, and continuous airflow ensures the removal of air pollutants and provides fresh air. Desk 2 is placed in the middle of the office where both artificial and natural lights contribute to the lighting levels, not directly impacted by the airflow from the ceiling diffuser to the exhaust. Desk 3 is located close to the ribbon window where lighting conditions are mainly driven by natural daylighting, outdoor heat gains are high and placed under the ceiling diffuser. Assessing sensor positioning across different locations in the room characterized by variations in lighting levels, thermal conditions, and airflow patterns helps to provide generalizable insights drawn regardless of desk placement in the room.

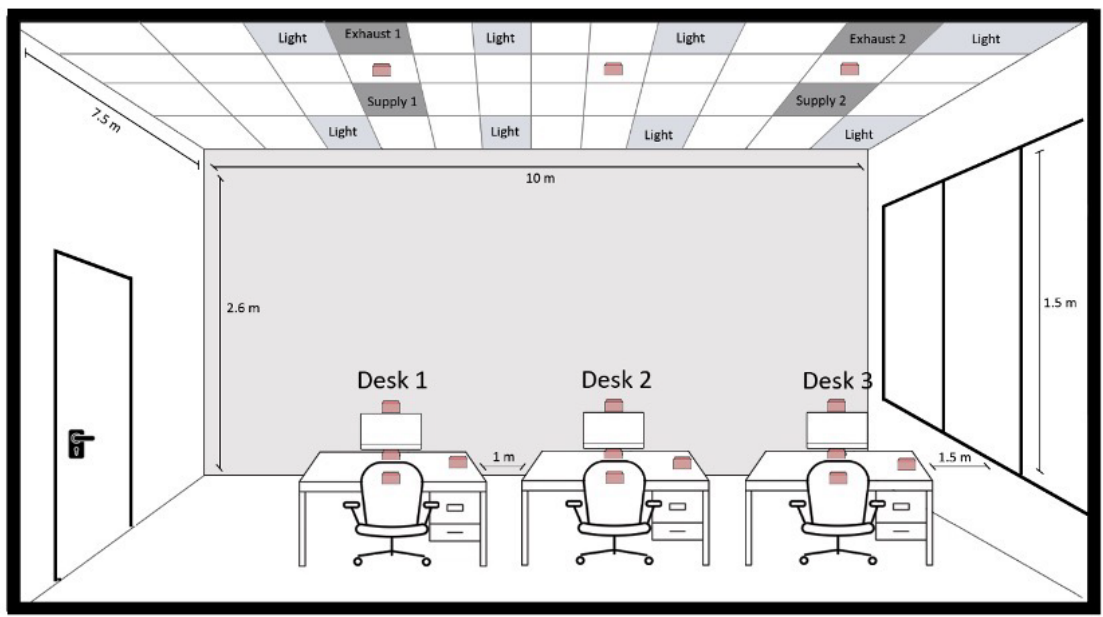

Figure 2. Living laboratory (open-plan office) configuration and characteristics.

## 2.2 Development of the IEQ sensor package

This paper presents a sensor package designed and assembled to monitor 4 IEQ parameters, comprising 8 metrics: IAQ (Carbon dioxide and Particulate matter 1, 2.5 and 10), visual (Horizontal illuminance), thermal (Temperature and Humidity), and acoustics (Sound) (Figure 3).

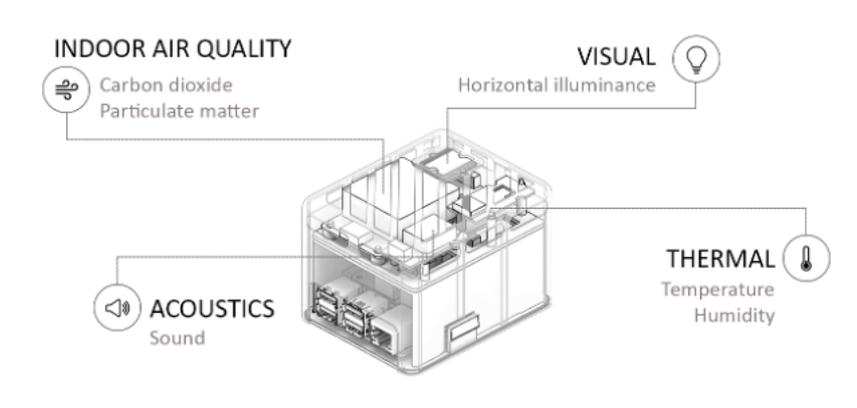

Figure 3. IEQ sensor package monitoring IAQ, visual, thermal and acoustic metrics.

### 2.2.1 Hardware and sensor package design

5 off-the-shelf sensors (brand: Seed Studio Grove) designed to detect the 8 IAQ metrics are secured to a 3D printed base shield and connected to the I2C ports of the base hat via standard wiring. Table 2 breakdowns the utilized sensors highlighting their measurement range and accuracy. The temperature and humidity sensor DHT22 combines capacitive humidity sensing and a thermistor-based temperature sensing. The Sensirion SGP30 is a gas sensor detecting $CO_2$ using Metal Oxide Semiconductor (MOx) technology. It features heated sensing elements that change resistance when specific gases interact, with resistance changes proportional to gas concentration, allowing electrical measurement and gas concentration determination. The Laser PM sensor HM3301 operates based on light scattering principles through a laser diode and a photodetector to quantify the presence of particles in the air. The Digital Light sensor TSL2561 measures the lux ambient level through a photodiode which converts changes in light intensity into electrical signals. Finally, the Sound Sensor is a module that detects sound levels in its surroundings by using a microphone element to pick up sound waves followed by the conversion of the sound waves into corresponding voltage variations.

Table 2. List of utilized sensors including measurements range, accuracy and resolution.

| Sensor | Parameter | Range | Accuracy | Resolution |
|---|---|---|---|---|
| **DHT22** | Temperature | -40 – 80 °C | ±0.5°C | 0.1°C |
| | Humidity | 5 – 99% RH | ±2% RH | 0.1% RH |
| **Sensirion SGP30** | $CO_2$ | 400 – 60.000 ppm | ±5% of reading | 1 ppm |
| **Laser HM3301** | $PM_1$ | 1 – 500 µg/m³ | ±0.3 µg/m³ | 1 µg/m³ |
| | $PM_{2.5}$ | | | |
| | $PM_{10}$ | | | |
| **Digital TSL2561** | Illuminance | 0.1 – 40.000 lux | ±15% lux | 0.1 lux |
| **Sound sensor** | Sound | 0 – 120 dB | ±0.5 dB | 0.1 dB |

A 3D printed enclosure was designed following a "box stacking" concept to ensure space saving and a slim design while fitting the sensors, wiring and the microprocessor. The boltless design allows easy assembly and maintenance. The upper half of the sensing device is based on an open concept to allow accurate measurement detection with no obstruction while facilitating airflow (Figure 4). Upon connection to the local internet, the Internet Service Provider (ISP) assigns a unique static IP address to the multi-sensing device. The continuous data is subsequently received by a computer and stored in a CSV file.

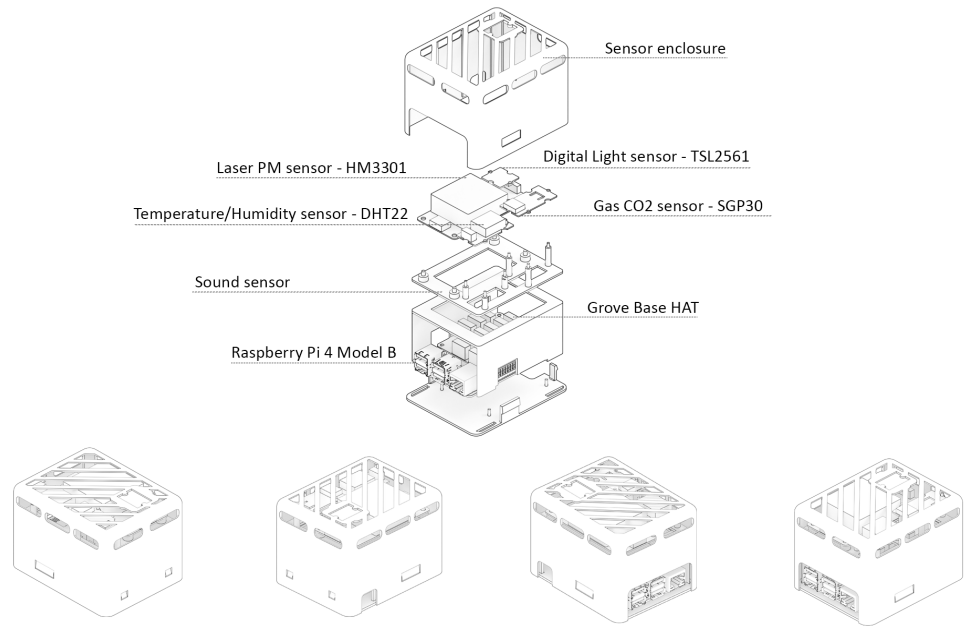

Figure 4. Exploded axonometric drawing of the components of the sensor package and enclosure box.

## 2.2.2 Validation analysis of the sensor package

A two-fold analysis was carried out to 1) validate the accuracy of the sensor package and 2) assess its compliance with standards. Initially, the IEQ sensor package was validated by placing it side-by-side with the corresponding high-precision laboratory-grade reference instrument during point-in-time monitoring (Figure 5). All laboratory-grade sensors used in the study underwent rigorous calibration to ensure accuracy. The Q-Trak 7575 was employed due to its established use in field applications and compliance with industry standards, including ASHRAE 55 (ASHRAE, 2017) and 62.1 (ASHRAE, 2022). While the Q-Trak is not the most advanced laboratory-grade device available, its balance between accuracy and practical deployment makes it suitable for real-world office monitoring scenarios. Parkinson et al. (2019b) have also demonstrated its reliability in similar IEQ studies. It was previously calibrated with traceable gas standards under controlled conditions, adhering to TSI guidelines. The device was exposed to controlled environments with known concentrations, and its sensor output was adjusted to match the reference standards. The DustTrak II 8532 was selected for its ability to continuously monitor PM concentrations across multiple size ranges ($PM_1$, $PM_{2.5}$, and $PM_{10}$). While potential biases in particle size distribution and optical limitations are noted (Li et al., 2019), its capacity to provide real-time data makes it valuable in dynamic environments, such as offices. It was calibrated against a gravimetric reference method with generated aerosol concentrations through a particle-dispersion machine. A calibration curve was developed to correct bias in the DustTrak readings, particularly to account for the overestimation of particle mass concentrations. Despite its limitations, DustTrak offers insights into PM level variations, aligning with the study's goal. Its real-time capabilities were especially useful during controlled experiments simulating human activities such as sneezing and coughing. The Testo light meter calibration was conducted using a traceable reference light source in accordance with ISO/CIE photometric standards. The output was verified against the reference values across a range of illuminance levels, and correction factors were applied where deviations were observed. The NTi XL2 Audio and Acoustic Analyzer was calibrated using a Class 1 acoustic calibrator certified to IEC 60942 standards. A reference sound pressure level of 94 dB at a frequency of 1 kHz was used to verify the instrument's sensitivity and accuracy. Additionally, field calibration checks were performed periodically during the study to ensure reliable measurements in varying environmental conditions.

The developed sensor package was cross-calibrated against the aforementioned sensors as readings where benchmarked under controlled conditions. These comparisons enabled the derivation of correction factors for the developed sensor package, which were applied during the study to enhance the reliability of its readings. The comprehensive calibration procedures ensured that all instruments, including both commercial-grade and developed sensors, provided accurate and reliable data throughout the study.

A 1-minute sample of concurrent measurements for all the IEQ metrics was collected every 30 mins from 10:00 to 18:00 to capture fluctuations of environmental parameters due to varying occupancy, lighting, sound, and thermal conditions. Measurements were repeated for 5 working days, following the same measurement protocol. This amounts to 80 point-in-time measurements collected. Table 3 breakdowns the reference instruments employed for the validation.

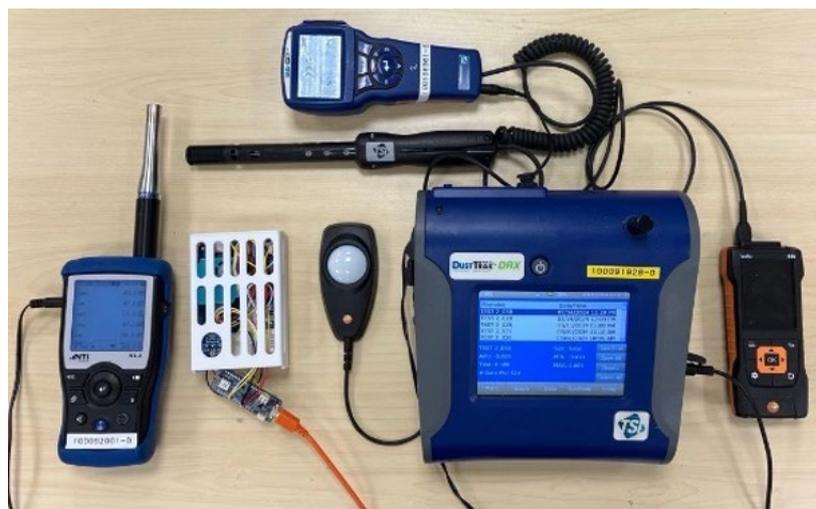

Figure 5. Validation setup of the IEQ sensor package with reference instruments.

Table 3. List of sensors included in the package and corresponding reference instrument for validation.

| Parameter | Low-cost sensor | Laboratory-grade instrument |
|---|---|---|
| **Temperature** | DHT22 | TSI Q-Trak 7575 |
| **Humidity** | DHT22 | TSI Q-Trak 7575 |
| **$CO_2$** | Sensirion SGP30 | TSI Q-Trak 7575 |
| **$PM_1$, $PM_{2.5}$, $PM_{10}$** | Laser PM sensor HM3301 | DustTrak™ II Aerosol Monitor 8532 |
| **Illuminance** | Digital Light sensor TSL2561 | Testo light meter |
| **Sound** | Sound Sensor | NTi XL2 Audio and Acoustic Analyzer |

In addition, longitudinal data were collected during a 2-week period of continuous monitoring by co-locating the IEQ sensor package with an off-the-shelf IEQ monitoring unit (Atmocube from Atmotech Inc.). Data collected from both the IEQ sensor package, and each reference equipment was used to calculate the standard error of the estimate (SEE) as a measure of equipment accuracy. While the coefficient of determination ($R^2$) is frequently utilized in linear regression to evaluate instrument accuracy, it is not suitable as an uncertainty measure for calibrated devices since it does not reflect the prediction error. Conversely, the SEE measures the accuracy of predictions made by the regression model and provides an absolute measure of fit reflected in the same units as the response variable (Parkinson et al., 2019b).

The results of the error analyses are summarized in Table 4, where the average SEE of the IEQ sensor package is presented for each parameter during point-in-time and continuous monitoring validation tests. The table also collates instrumentation requirements and thresholds for accuracy from standards for measurement in commercial offices across the four categories of IEQ.

Table 4. Summary of SEE of IEQ sensor package during both point-in-time and continuous monitoring and corresponding threshold in the prominent standards.

| IEQ category | Parameter | SEE Point-in-time | SEE Continuous | Threshold | Source |
|---|---|---|---|---|---|
| Thermal | Temperature | 0.44 °C | 0.24 °C | 0.50 °C | ISO 7726:2001 |
| | Humidity | 2.86 % | 1.47 % | ± 5% | ASHRAE 55-2017 |
| Indoor Air Quality | $CO_2$ | 28.12 ppm | 10.25 ppm | 50 ppm | ASHRAE 62.1/ 62.2 |
| | $PM_1$ | 1.55 µg/m³ | 1.27 µg/m³ | 2 µg/m³ | WELL ® |
| | $PM_{2.5}$ | 1.87 µg/m³ | 1.63 µg/m³ | 2 µg/m³ | WELL ® |
| | $PM_{10}$ | 1.86 µg/m³ | 1.71 µg/m³ | 2 µg/m³ | WELL ® |
| Lighting | Illuminance | 21.11 lux | 10.52 lux | 5% range | ISO/CIE 19476 |
| Acoustics | Sound | 0.92 dB | 0.43 dB | ±1 -1.1 dB | ANSI S1.43 / EN 61672-1 |

The results obtained from both the point-in-time and continuous monitoring are further evaluated to validate the compliance of the sensor package with relevant standards for IEQ measurement, summarized by Parkinson et al. (2019b). While performance measurements suggest that SEE values of IEQ sensor package against high-precision instruments obtained from point-in-time monitoring are higher compared to those recorded during continuous monitoring, all SEE values are within the maximum allowed thresholds and complying with prominent standards. The SEE of temperature (0.44 °C - 0.24 °C) and humidity (1.47 % - 2.86 %) data collected from the sensor package are within the tolerance range of 0.5°C as recommended by the standard ISO 7726:2001 (ISO Standard 7726, 2001), and ± 5% according to ASHRAE 55-2017 (ASHRAE, 2017). Similarly, with a calculated SEE value of 10.25 ppm and 28.18 ppm, $CO_2$ values comply with the threshold of 50 ppm + 3% of reading at values between 400 and 2000 ppm derived from WELL Building Standard (V.2 Well, 2018). In addition, SEE values for PM (between 1.27 µg/m³ and 1.87 µg/m³) are within the maximum allowed threshold of 2 µg/m³ + 15% of reading at values between 0 and 150 µg/m³, as per WELL Building Standard.

In addition to other parameters, illuminance uncertainty was compared with the ISO/CIE 19476 standard, which sets a threshold of 5% of the reading. The calculated SEE for illuminance during point-in-time measurements was 21.11 lux, corresponding to approximately 3.9% of the mean reading (540 lux), and for continuous monitoring, the SEE was 10.52 lux, corresponding to approximately 1.9% of the mean. Both values are well within the ISO/CIE 19476 threshold, further confirming the sensor package's compliance with industry standards for lighting measurements. In conclusion, The SEE of acoustic sensor (0.43 dB and 0.93 dB) is within the tolerance range of ±1 dB for general field use according to standard ANSI S1.43 (ANSI S1.43, 1997) and ± 1.1 dB based on EN 61672-1 (IEC, 2013). In conclusion, the error analyses indicate the adequacy of the sensor package to monitor IEQ conditions and detect parameters' trends and compliancy over time. This represents the desired purpose of a continuous monitoring system.

### 2.3 Sensor placement for personalized monitoring

The positions of the sensor package in this paper were informed by current best practices but tailored to investigate personalized IEQ, integrated within the immediate workplace environment (Coleman and Meggers, 2018; Kierat et al., 2022; Pantelic et al., 2020; Pei et al., 2019). 15 stationary IEQ sensing packages (IDs 1-15) are developed and placed in 5 monitoring locations around the occupant at 3 desks to collect concurrent measurements of temperature, humidity, $CO_2$, $PM_1$, $PM_{2.5}$, $PM_{10}$, illuminance and sound with a sampling interval of 5 seconds (Figure 6). Table 5 details the conventional naming adopted henceforth and their positioning around the occupant. Sensing devices 1, 9 and 13 were placed in the middle of the desk, directly below the computer screen at a height of 0.7 m and 0.5 m from the occupant. Sensor packages 2, 10 and 12 were positioned on top of the computer screen at a height of 1.20 m and 0.7 m from the occupant. Sensor packages 4, 8 and 11 were located on the right side of the desk at a height of 0.7 m and 0.9 m from the occupant. Sensor packages 3, 7 and 13 were positioned on the ceiling tiles right above the occupant at a height of 2.6m and 1.4 m from the occupant. Finally, sensor packages 5, 6 and 15 were hooked to the backrest of the chair at a height of 0.7 m and 0.4 m from the occupant.

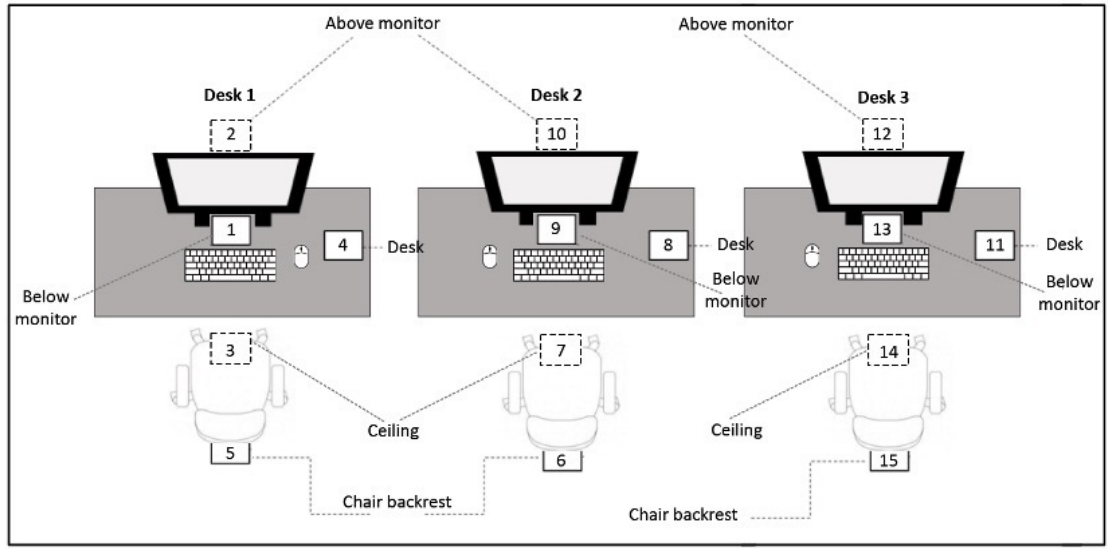

Figure 6. Personalized sensor positioning across desks and ID numbers.

Table 5. Sensor package naming and placement.

| Sensor package name (IDs) | Sensor package placement |
| --- | --- |
| Below monitor (1, 9, and 13) | 0.7 m height, 0.5 m from occupant |
| Above monitor (2, 10, and 12) | 1.2 m height, 0.7 m from occupant |
| Desk right side (4, 8, and 11) | 0.7 m height, 0.9 m from occupant |
| Ceiling (3, 7, and 14) | 2.6 m height, 1.4 m from occupant |
| Chair backrest (5, 6, and 15) | 0.7 m height, 0.4 m from occupant |

## 2.4 Experimental procedure

Two main experiments were conducted over two time periods (02/12/2023 – 02/01/2023; and on 04/02/2024), a longitudinal experiment with occupants and a short-term experiment utilizing a particle dispersion machine (Figure 7). These consisted of a set of experimental cases, as detailed in Table 5. Initially, a 2-week longitudinal study of concurrent monitoring of the 3 desks was performed. As depicted in Table 6, 2 sensor package orientations were tested during the two experimental periods: horizontal orientation where the sensors face the ceiling (first week) and vertical orientation where the sensors directly face the occupant (second week). Two healthy males and one female represented the occupants, and the same composition was kept the same for each scenario. The age of the occupants was between 21 and 30, with BMI ranging between 18.2 kg/m$^2$ and 23.5 kg/m$^2$. During the experiments, the occupants wore typical summer office clothing (average 0.5 Clo). Although some variables such as temperature, humidity and lighting level may remain stable in a well-mixed environment, the primary goal of this study is to investigate variations in local conditions that occupants directly experience. Sensors were positioned in close proximity to the occupants to capture micro-environmental changes that might still occur due to factors such as metabolic heat generation, clothing variations, or the presence of electronic devices, which can influence localized thermal comfort and airflows. The inclusion of three human subjects allowed to examine the micro-environments surrounding each individual, as these could vary during the working day.

In addition, to introduce controlled perturbations and gain a deeper understanding of the effects of sensor positioning, a series of short-term experiments were conducted to identify sensor positioning with respect to pollutants produced by occupants. Two scenarios recreated events that commonly occur in office buildings and increase the air pollutant concentrations: 1) human sneeze event, and 2) human cough event. These perturbations generated localized variations in air quality parameters, particularly PMs, allowing to evaluate how sensor positioning influences the detection of such changes near the occupants. Each experimental scenario recreating a real-life event lasted 10 minutes and involved the sequential monitoring of 2 desks (desk 1 placed under the ceiling diffuser where continuous airflow ensures the removal of air pollutants and desk 2 not directly impacted by the airflow from the ceiling diffuser to the exhaust) and explored two sensor package orientations (facing the ceiling and the reference occupant).

A particle dispersion machine developed by the Hong Kong University of Science and Technology, Research and Design Corporation Limited, simulated the real-life events of a person sneezing and coughing. The machine generated droplets and aerosols with a mean diameter of 13.5 μm (cough) and 56 μm (sneeze) (Chao et al., 2009) and velocities of 11.2 m/s (cough) and 14 m/s (sneeze) (Zhu et al., 2006) using a mixture of glycerol (10%) and water (90%) to resemble the properties of human saliva and its evaporation. The particle dispersion machine was placed where an occupant would typically sit, adjusted to a height of 1.1 m above the ground. The air-atomizing nozzle expelled droplets and aerosols horizontally toward the desk to simulate real-life dispersion events.

To determine the ground truth data, the calibrated DustTrak sensor was positioned directly adjacent to the nozzle of the particle dispersion machine (Figure 8). The sensor's probe was connected to the nozzle using a short, flexible tube that minimized particle loss and ensured that measurements were taken immediately after particle ejection. While the particle dispersion machine was calibrated to produce droplets of a known size and velocity, the concentration of particles was not fixed. Instead, the reference sensor measured the real-time concentration of particles ejected during each dispersion event, thereby providing a reliable benchmark for evaluating the performance of other sensor placements.

In this study, the term "ground truth" specifically refers to the data collected by this reference PM sensor. It provided accurate measurements of particle concentrations directly at the point of emission, where the characteristics of the particles—such as size and velocity—were known and controlled. The use of this setup ensured a consistent and reliable benchmark for evaluating sensor accuracy under controlled experimental conditions. Each dispersion event (cough or sneeze) lasted for 0.55 seconds, with an interval of approximately 1 minute between each event to allow for sufficient particle dispersion. Following the completion of each experimental scenario, the open-plan office was sealed for 30 minutes to monitor the decay of air pollutant concentrations and ensure that background levels were sufficiently low before the subsequent experiment. The total duration of the experiment was 7.5 hours, during which various sensor placements were systematically evaluated against the reference sensor. By combining steady-state monitoring with controlled perturbations, the study aimed to provide a comprehensive understanding of how sensor placement impacts the capture of both stable and transient environmental variations.

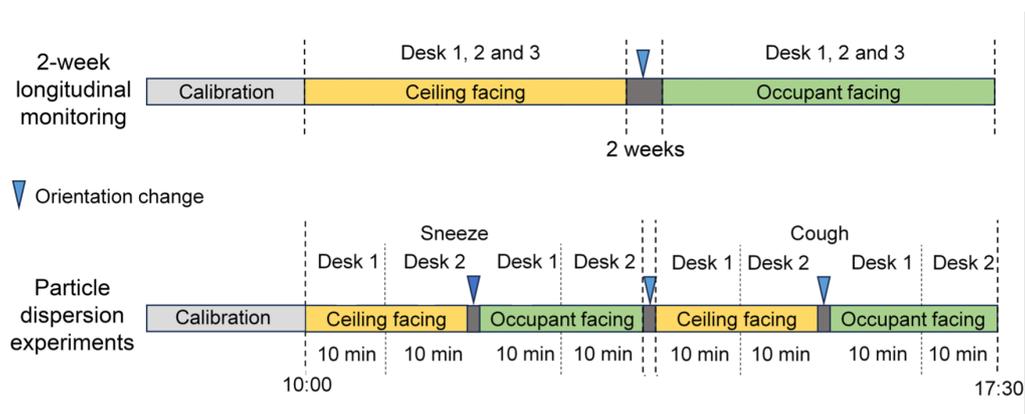

Figure 7. Timelines of the 2-week longitudinal monitoring and particle dispersion experiments.

Table 6. Experimental cases to investigate sensor positions for personalized monitoring of IEQ.

| Experiment | Desk layout | Sensor package orientation | Sensor package position and IDs |
|---|---|---|---|
| Long-term (3 human subjects) | 3 desks 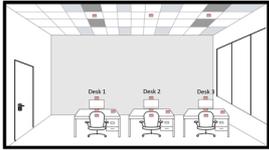 | Ceiling/Occupant 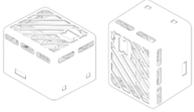 | 5 positions (IDs 1 to 15) 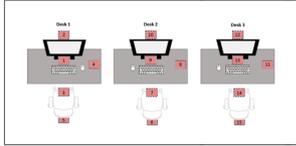 |
| Short-term (sneeze + cough) | 2 desks 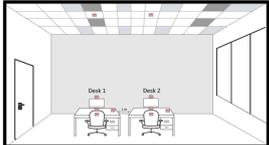 | Ceiling/Occupant 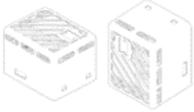 | 5 positions (IDs 1 to 10) 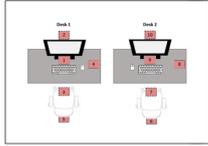 |

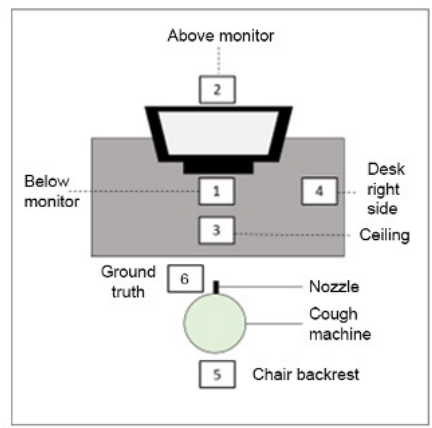

Figure 8. Particle dispersion machine location with respect to the sensor positions.

## 2.5 Data analytics

Due to the impracticality of obtaining ground-truth data for a wide variety of IEQ parameters in the context of sensor placement, a dimensionality reduction technique to transform and simplify high-dimensional data while preserving essential information (PCA) was employed to identify the recommended position of a sensor from the longitudinal data collected across multiple locations. This position contributes the most to the variance in measurements, capturing the maximum information in relation to the occupant. This study aims to identify the positioning of a sensor package that combines the 4 main IEQ parameters (thermal, visual, acoustic, and IAQ), and for each individual IEQ metric (e.g., PMs) and grouped IEQ parameters (e.g., IAQ).

To ensure that each metric contributes equally to the analysis, the data collected at different positions are standardized to have a mean of 0 and a standard deviation of 1. Let X be the matrix containing the sensor readings across the 5 positions, where each row represents a data sample, and each column represents a sensor position (Equation 1).

$$X_{standardized} = \frac{X - mean(X)}{std\ (X)}$$

(1)

Subsequently, the covariance matrix (C) of the standardized data is computed. This matrix quantifies the relationships and correlations between the different variables collected across the five positions (Equation 2). Where n is the number of data samples.

$$C = \frac{1}{n} X^T_{standardzed} X_{standardized}$$

(2)

The eigenvalues (λ) and eigenvectors (*v*) of the covariance matrix (C) are then calculated (Equation 3). The eigenvectors indicate the directions of maximum variability in the data, while the eigenvalues represent the amount of variance along each eigenvector.

$$C_v = \lambda_v$$

(3)

The eigenvectors are sorted in descending order based on their corresponding eigenvalues. The eigenvector with the highest eigenvalue corresponds to the first principal component (PC), these are the new variables that capture the most significant variance in the data. Subsequent eigenvectors capture progressively less variability. The original data from the different sensor positions are then projected onto the principal components to obtain the principal component scores ($PC_{score}$). This projection transforms the data into a new coordinate system aligned with the directions of maximum variability (Equation 4). Where V is the matrix containing the selected eigenvectors.

$$PC_{score} = X_{standardized} \times V$$

(4)

By examining the coefficients of the projected data along each principal component, it is possible to determine which position contributes the most to the overall variance in the dataset. The position that contributes the most to the highest principal component can be considered the recommended sensor position, as it captures the most variations in the data.

To identify the sensor position that best represents personal exposures to the investigated air pollutants, this study calculated Spearman's rank correlation analysis (ρ) to assess the strength and direction of the monotonic relationship between the ground truth measurements collected from the reference sensor and each sensor position. The resulting correlation coefficient ρ (rho) ranges between -1 and 1. A value of +1 indicates a perfect positive monotonic correlation, -1 indicates a perfect negative monotonic correlation, and 0 indicates no monotonic relationship between the variables. Similarly, $R^2$ and Mean Absolute Error (MAE) between the measurement of the reference sensor and each sensor located in different positions were calculated to quantify the deviations between the two sets of measurements. Lower error values indicate better agreement between the sensors. The strength of correlation, $R^2$ and magnitude of error for each sensor position are then ranked and compared. The sensor location that most accurately reflects the environmental conditions experienced by occupants exhibits the strongest correlation, and the lowest error compared to the reference sensor, indicating a better match in measurements.

## 3. Results

This section presents the results obtained from the experimental study on IEQ sensor placement in office environments. The findings are divided into four main parts:

- Section 3.1 provides a descriptive analysis of the IEQ metrics collected from different sensor positions and orientations. The goal is to examine how sensor placement affects the measurements of these metrics in various office conditions.

- Section 3.2 focuses on identifying the spatial positioning of sensors for each individual IEQ metric.

- Section 3.3 and 3.4 extend the analysis to grouped IEQ parameters, such as thermal, visual, acoustic, and air quality metrics, and combined all-in-one IEQ, aiming to determine the positions for capturing multiple IEQ metrics simultaneously.

- Section 3.5 discusses the results of short-term controlled experiments simulating real-life pollution events (e.g., coughing, and sneezing). This section assesses the impact of sensor placement on the detection of air pollutants like PMs.

### 3.1 Sensor Uncertainty Analysis

To evaluate the reliability of the sensor measurements and their influence on the results presented, an uncertainty analysis was conducted. This analysis considers the mean and standard deviation in individual sensor readings across the investigated IEQ parameters, quantifies the standard error of the mean (SEM) and computes the confidence intervals for a 95% confidence level (Figure 9).

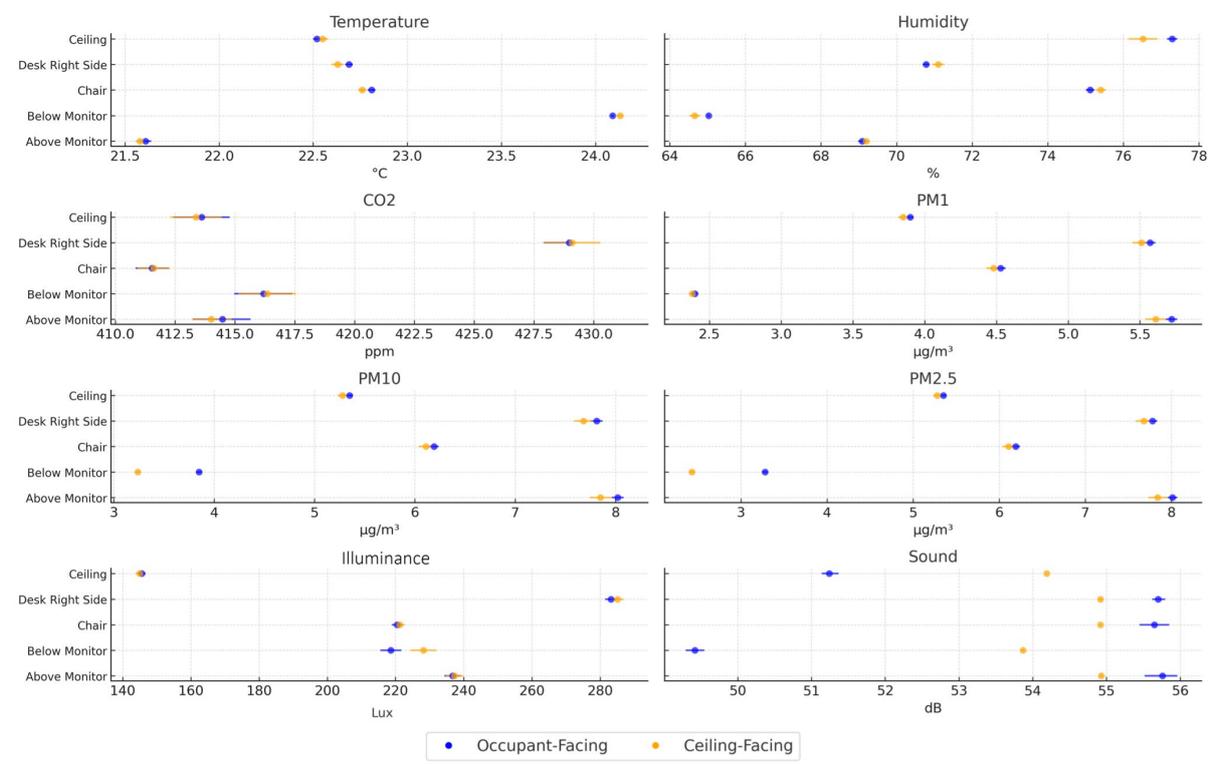

Figure 9. Uncertainty analysis of sensor reading measurements for each IEQ parameter.

For thermal comfort metrics, temperature readings exhibited deviations consistently within ±0.1°C, and relative humidity deviations remained within ±1%, both of which are minor compared to the observed ranges (21.5–24.0°C and 64–78%, respectively). IAQ metrics also demonstrated minimal uncertainty. For instance, $PM_1$, $PM_{2.5}$, and $PM_{10}$ measurements showed deviations of less than ±0.2 µg/m³, ensuring that variability is small relative to the measured concentrations, which spanned 3–8 µg/m³. Similarly, $CO_2$ concentration displayed deviations within ±3 ppm, an uncertainty that is negligible compared to the absolute range of 410–430 ppm.

Visual comfort and acoustic comfort metrics also reflected minimal variability. Illuminance readings showed deviations of no more than ±10 lux, representing less than 5% of the measured values (140–280 lux). Sound level measurements, ranging from 50 to 56 dB, exhibited standard deviations no greater than ±0.5 dB. These deviations are small compared to the magnitude of the recorded values, affirming the precision of the measurements.

These results demonstrate that the degree of uncertainty in the sensor readings is minimal relative to the observed magnitudes of the measured parameters. Consequently, the propagation of sensor error in subsequent calculations, including PCA, is unlikely to result in significant inaccuracies. The small deviations observed validate the robustness of the study's methodology and confirm the reliability of the findings.

## 3.2 Spatial positioning of sensors for individual IEQ metrics

Figure 10 depicts the eigenvalues of the first principal component of the PCA ranked according to their variance to identify the spatial positioning of sensors for each individual IEQ metric. Generally, it appears that different positions were identified for different IEQ metrics. Additionally, it is noticeable that positions vary depending on the desk location in the monitored office and the orientation of the sensors. However, sensors oriented facing the occupants generally display better agreement across different desks. Temperature should be measured below the computer screen at a distance of 0.5 m from the occupant and the sensor should be placed horizontally on the table, facing the ceiling, as it displays the highest loading across desks. This position aligns with the "breathing zone" and provides a representative measurement of the thermal conditions that occupants typically experience while working at their desks. In addition, the ceiling-facing oriented ensures that the sensor measures the ambient air temperature rather than being influenced by localized heat emissions such as computers or equipment on the desk. Comparatively, the recommended position for the humidity sensor is above the monitor, facing the reference occupant at a distance of 0.7 m. Placing the humidity sensor at eye level, allows for measurements that better reflect the humidity conditions that occupants experience and where they are most likely to experience it and be affected by it, while minimizing the likelihood of localized disturbances from the environment.

$CO_2$ could be bundled with the temperature sensor and positioned horizontally on the table below the computer monitor, since this placement obtained the highest loading across the desks. Similarly, being positioned within the occupants' exhalation area, this location ensures that the $CO_2$ measurements captured are relevant to their immediate environment.

The sensor capturing PM values exhibits similar patterns for the three investigated sizes of particles ($PM_1$, $PM_{2.5}$, and $PM_{10}$) and their recommended positioning is above the monitor, facing the reference occupant, as identified for humidity. In certain indoor environments, particle behaviour and dynamics may not vary significantly across different size fractions. This could be due to several factors such as: homogeneity and uniformity of particle sources, effective spatial mixing, and distribution due to the ventilation and airflow patterns or the similar sampling mechanism of the sensors that make their measurements respond similarly to changes in particle concentrations.

Illuminance values should be collected from a sensor located on the right side of the desk, 0.9 m from the occupant. This represents the closest position to the window and helps to assess the impact of daylighting on illuminance levels at the workspace. Furthermore, this is positioned where task lighting, such as a desk lamp, is commonly used. Therefore, measuring illuminance levels in this location can investigate the combined effect of natural and artificial lighting. In addition, this position offers the highest accuracy as it is unaffected by secondary lighting sources such as the computer monitor. Finally, the position that better captures sound is below the computer screen, directly facing the occupant, similarly to temperature and $CO_2$, which could be packaged within a single sensing device. This allows for the assessment of sound levels in close proximity to the main sound sources that impact the acoustic quality of the workspace since work tasks often involve activities like typing, phone calls, or video conferences. Placing the sensor facing the occupant evaluates noise levels during these tasks, providing information on sound levels that may affect occupant performance, capturing the conditions that are truly experienced by the occupant at the workstation.

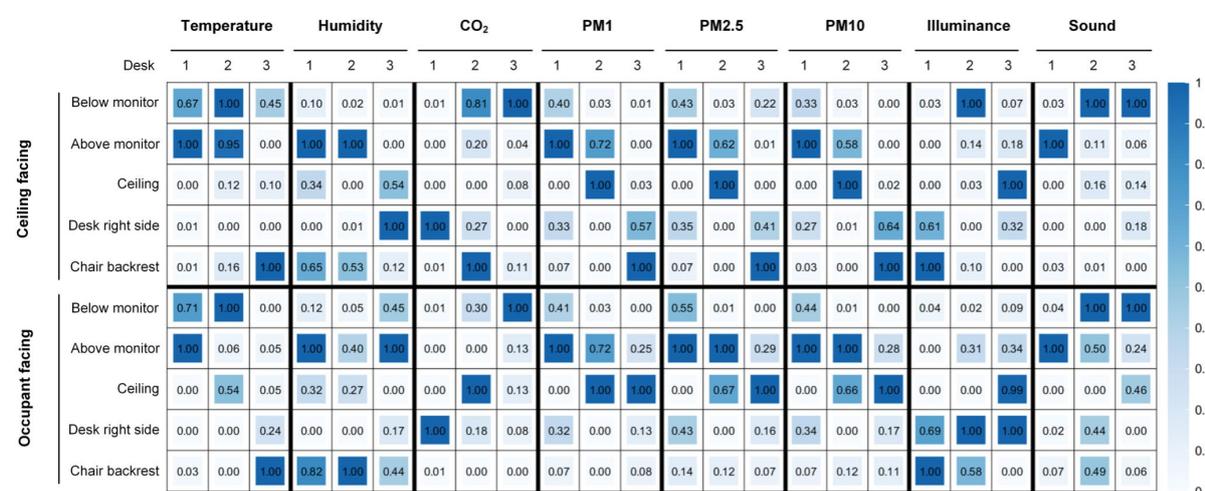

Figure 10. Recommended positioning of sensors for individual IEQ metrics.

## 3.3 Spatial positioning of sensors for grouped IEQ parameters

The eigenvalues of the matrix that merges the individual metrics pertaining to their respective IEQ parameters are presented in Figure 11, ranked according to their loadings. In general, different positions were identified for different IEQ parameters. However, across all monitored desk locations, the sensor package facing the occupant consistently captured IEQ conditions as experienced by the occupant in the workstation. The sensor measuring thermal parameters (temperature and humidity) should be placed on the chair backrest, at a height of 0.7 m and 0.4 m from occupant. This represents the closest location to the reference occupant. In fact, the chair backrest is a key point of contact between the occupant and the workspace. Therefore, placing the sensor in this position can measure thermal conditions in the immediate vicinity of the occupant's body, which is particularly relevant for evaluating thermal comfort. In addition, occupant body can influence local thermal conditions. Placing the sensor on the chair backrest, near the occupant body, captures the effects of their presence/absence on temperature and humidity levels.

Comparatively, the highest loading across positions for the IAQ parameters ($CO_2$, $PM_1$, $PM_{2.5}$, $PM_{10}$) was identified from the sensor above the monitor as this corresponds to the position closer to the breathing zone of the occupant (1.2 m height from the ground and 0.7 m from the occupant). The distance between the IAQ sensor and the occupant affects the accuracy of the inhalation exposure detection. Placing IAQ sensors above the monitor provides a generalized assessment of IAQ in the workspace. Therefore, positioned at or near the occupant's head height, this position captures a combination of factors that influence IAQ and offer insights on the quality of air they breathe, including ventilation effectiveness, pollutant distribution, and occupant exposure, as they are experienced by occupants. As detailed in Section 4.1, illuminance should instead be measured on the right side of the desk, as this is the position closer to the window, considered the main source of lighting. Finally, the position below the computer screen appears to better capture sound, as it is the closest to the occupant (0.5 m), without obstructions, and can capture the noise level derived from work tasks.

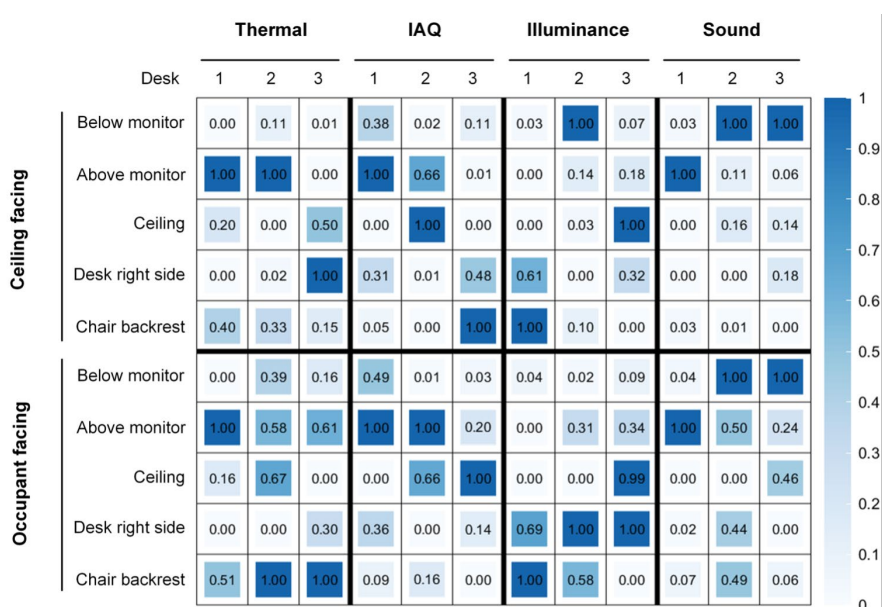

Figure 11. Recommended positioning of sensor package for grouped IEQ parameters.

## 3.4 Spatial positioning of sensors for combined IEQ parameters

The loadings of the matrix that combines all the 4 IEQ parameters are illustrated in Figure 12, from which the recommended positioning of the sensor package is derived. While positioning the sensors package horizontally, facing the ceiling, different positions are observed based on the desk location. However, it is noteworthy that when the sensor package is oriented to face the occupant, the position above the monitor consistently captures IEQ conditions as experienced by the occupant in the workstation, regardless of the room desk layout. This represents the sensor location which strikes a balance across all the IEQ metrics simultaneously. This position ensures that the data collected reflects conditions as they are experienced by the occupants of the workspace.

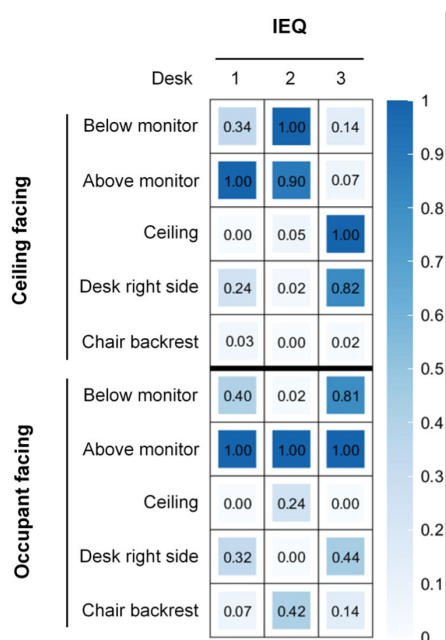

Figure 12. Recommended positioning of sensor package for combined IEQ parameters.

## 3.5 Simulated cough and sneeze experiment (PM1, PM2.5, and PM10)

It is notable that sensors positioned on the chair backrest and ceiling do not align with the measurement trends obtained by the ground truth sensor, regardless of sensor orientation (Figure 13). Comparatively, sensors placed below and above the monitor, as well as on the right side of the desk can detect particles from both coughing and sneezing events, although characterized by different strengths of correlation, $R^2$, and MAE values. Furthermore, the intensity of the sneezing events resulted in higher PM values being recorded compared to simulated coughing events.

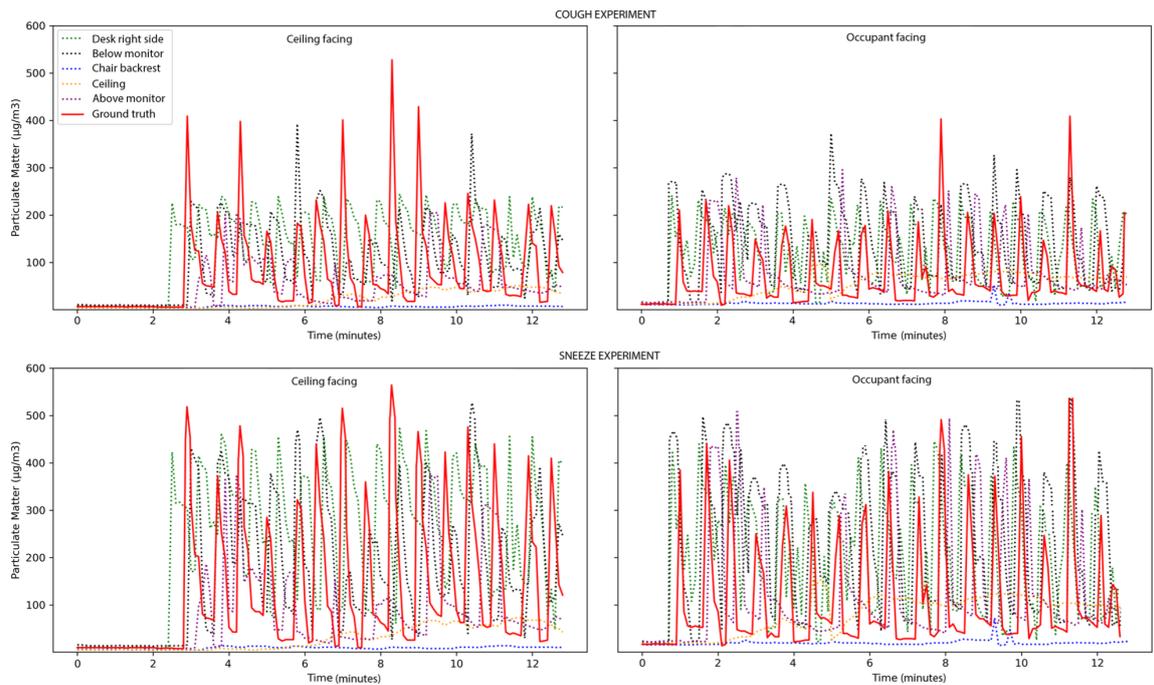

Figure 13. Time series across sensor positions and ground truth data for increased PMs due to coughing and sneezing.

The correlation, $R^2$ and MAE patterns between each sensor location and the reference sensor (ground truth) positioned close to the nozzle of the particle dispersion machine are similar for $PM_1$, $PM_{2.5}$, and $PM_{10}$ (Figure 14). In fact, in some indoor environments, the sources of PM, regardless of their size fractions, may be relatively uniform and evenly distributed. Furthermore, distinctions between the size fractions can be blurred and their behaviour might be similar regardless of size if smaller particles tend to agglomerate into larger particles or if the particles are primarily generated from a single source and have similar settling velocities. However, values vary across $PM_1$, $PM_{2.5}$, and $PM_{10}$ since particles may behave differently in the air and this behaviour could affect how they are sampled and measured. For example, smaller particles (PM1) tend to stay suspended longer, while larger particles (PM10) may settle quicker.

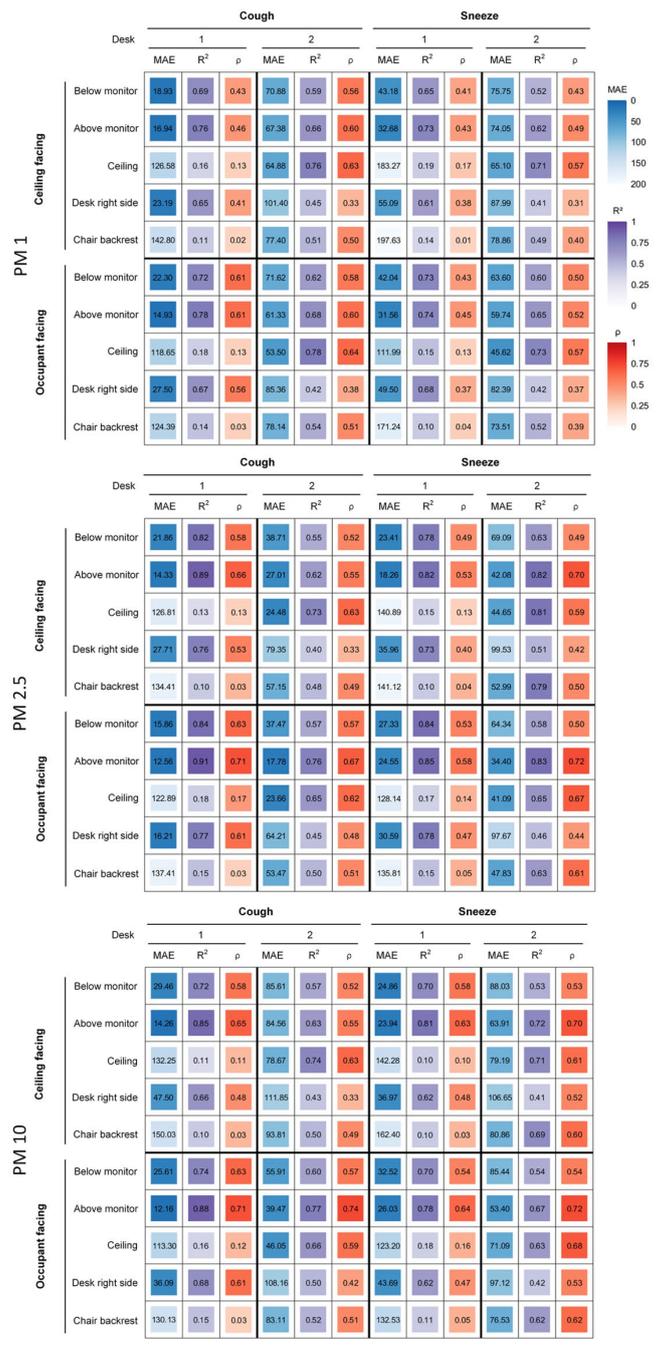

Figure 14. Correlation and error metric between each position and reference sensor during the simulated cough and sneeze experiment for $PM_1$, $PM_{2.5}$, and $PM_{10}$.

In addition, the indoor environment's ventilation and airflow patterns can play a significant role in mixing and distributing particles of different sizes. During the coughing and sneezing events, on average across positions and PM sizes, the sensor oriented facing the occupant displays higher correlation values (coughing ρ: 0.41, sneezing ρ: 0.34) for the desk located underneath the diffuser while a higher correlation value is identified for the ceiling-facing sensor (coughing ρ: 0.48, sneezing ρ: 0.36), for the desk location not impacted by the airflow from the ceiling diffuser to the exhaust. The presence or absence of the ceiling diffuser can create different airflow patterns within the desk area. This can affect how cough-and sneeze-generated particles disperse and reach the sensor. In the desk location not impacted by the diffuser, the absence of direct airflow from the diffuser to the exhaust may result in more stable and less turbulent local airflow conditions. This could allow for more predictable and consistent dispersion of particles and potentially higher correlations with the ceiling-facing sensor.

Average MAE values for the PM particles detected from both coughing and sneezing events at the desk located underneath the diffuser are significantly higher (coughing: 157.5 µg/ $m^3$, sneezing: 164.7 µg/ $m^3$) compared to the desk away from the diffuser (coughing: 54.7 µg/ $m^3$, sneezing: 69.7 µg/ $m^3$). The turbulence in airflow conditions can affect how particles disperse and interact with the sensors, potentially resulting in higher MAE values. In addition, sensor orientation appears to have an impact on the average MAE values as placing the sensor facing the occupant displays a lower average MAE (coughing: 62.8 µg/ $m^3$, sneezing: 54.71 µg/ $m^3$) compared to the ceiling-facing oriented sensor (coughing: 79.3 µg/ $m^3$, sneezing: 70.6 µg/ $m^3$). In fact, the sensor facing the occupant is positioned to directly intercept particles released during coughing and sneezing events. This direct exposure allows the sensor to capture a more representative sample of the particles, leading to more accurate measurements and lower MAE values. In addition, the sensor facing the occupant has a shorter path for particles to travel before reaching the sensor. This shorter path reduces the potential for particle dispersion and dilution in the air, resulting in more precise measurements. Finally, sensors facing the ceiling may be more susceptible to interference from other sources, such as particles settling from the air or particles introduced by ventilation systems. These interferences can lead to higher MAE values as the sensor may detect particles that are not directly related to the coughing or sneezing event.

Specifically, the sensor positioned above the computer monitor and facing the occupant displays the highest average correlation with the reference sensor during both the coughing ($\rho$: 0.64) and sneezing ($\rho$: 0.56) events, regardless of desk location. This can be attributed to the direction of the particles sprayed and the proximity effect as these two sensors were the closest to each other on the spraying trajectory. It is interesting to note that the sensor above the monitor facing the occupant displays simultaneously the highest average $R^2$ values (coughing: 0.83, sneezing: 0.78) and lowest MAE (coughing: 21.96 µg/ $m^3$, sneezing: 34.57 µg/ $m^3$). These findings overlap with the results obtained from the PCA and enhance the reliability of the sensor above the monitor as most suitable position for PM.

## 4. Discussion

The results of this study highlight the importance of sensor positioning for effective personalized IEQ monitoring. The primary objective of many IEQ monitoring efforts is to assess and improve the comfort and well-being of occupants. However, current building standards, such as ASHRAE Standard 55 (ASHRAE, 2017) and 62.1 (ASHRAE, 2022), WELL v2 (V.2 Well, 2018), and RESET v2 (RESET® standard, 2022), provide only room-level guidelines for sensor placement. These standards do not offer detailed recommendations for sensor placement close to occupants, which limits their applicability in capturing the true experience of the occupant. Therefore, current practices are based on rule-of-thumbs or ease of installation (Yun and Licina, 2023).

The descriptive analysis in this paper demonstrated that even within a personal workstation space, the position and orientation of the sensor significantly affects data reliability monitoring of IEQ parameters, and the influence varies among different parameters. This study demonstrated that sensor positions for different IEQ metrics can vary spatially within an office indoor environment as different IEQ metrics are influenced by various factors and phenomena. In fact, different positions and orientations were identified for different IEQ metrics.

Bundling multiple IEQ sensors on the same device is increasingly emerging for continuous IEQ monitoring applications as it can reduce installation costs, sensors expenses, and power use. Nevertheless, this approach may introduce conflicts in positioning requirements due to varying sensor needs, necessitating compromises between placement for one sensor and avoiding inaccuracies in others. This paper found that placing the sensor package above the monitor allows to consistently measure conditions experienced by occupants in an office setting, regardless of their desk location. Furthermore, placing the sensor package above the computer monitor favours the integration into the workspace layout without interfering with desk activities. This is aesthetically pleasing and ergonomic, as it keeps the desk uncluttered and minimizes obstructions in the occupant's workspace while ensuring that IEQ measurements remain relevant. This is particularly important for IEQ parameters, where variations can lead to discomfort and health concerns. The occupant's immediate environment is where they are most exposed to various IEQ parameters. In addition, sensors facing the occupant are more likely to capture conditions as they are experienced by the occupants in the workstation, leading to better agreement across different desk locations in the room. Based on the recommendations derived from this study, building management and control systems can use sensor data to effectively adjust environmental conditions to maintain personalized comfort levels.

## 5. Limitations

The experiments were conducted across 3 desks located in 3 different locations of the living laboratory to cover different variations in lighting, thermal conditions, and airflow patterns and increase the generalizability of the recommendations. Although the desks selected for the experiments represent common scenarios occurring in office settings, the results are constrained to these locations, and it is uncertain whether different placements of the workstations would lead to different results. Additionally, although the explored sensor locations aim to cover a variety of personalized positions at different heights and distances from the occupant and they were abundant (5), the results are limited to the investigated positions. Furthermore, the experiments were conducted in a room where air-conditioning and ventilation was supplied through a VAV system, characterised by an east-facing 1.5m tall ribbon window spanning the entire length of the office. However, the layout, design, and configuration of the monitored office can significantly impact the distribution of environmental parameters. Therefore, it is unclear whether a different conditioning system or a different window layout would provide different conclusions. Hence, the insights are only valid for the aforementioned assumptions and the proposed positions might not be fully applicable to different office contexts and layouts, stationary sensor locations and room characteristics.

Further research could explore the reliability of the findings in different building types (e.g., classrooms), characterized by different occupancy dynamics, layouts, lighting, airflow patterns, and thermal conditions. Finally, although the aim of this paper is the identification of spatial positioning of a sensor package that combines the 4 main IEQ parameters (thermal, visual, acoustic, and IAQ), recommendations for the positioning of each individual IEQ metric and grouped IEQ parameters are also presented. However, it is acknowledged that they do not exhaustively encompass all potential combinations of sensor packages (e.g., packages including PMs, temperature and illuminance) and IEQ parameters. These combinations are beyond the scope of this paper as they are specific to individual use cases.

## 6. Conclusions

Improving the effective monitoring of IEQ is crucial to aid the shift towards occupant-centric buildings. However, there is a lack of recommendations on personalized placement and positioning of IEQ sensor packages for personalized monitoring in office settings. This study identified the most suitable positioning of a sensor package that monitors temperature, humidity, carbon dioxide ($CO_2$), particulate matter (PM), illuminance, and sound monitoring in an air-conditioned open-plan office characterized by dynamic occupancy.

Continuous IEQ data were gathered from 5 personalized sensor positions: above and below the computer monitor, on the right side of the desk, on the ceiling above the reference occupant, and attached to the backrest of the chair. These positions were explored under two different sensor orientations: horizontal, where the sensors faced the ceiling, and vertical, where the sensors directly faced the occupant. A longitudinal experiment spanning two weeks involved 3 occupants seated at 3 distinct desk locations, each characterized by varying thermal, lighting, and airflow conditions, were monitored to improve generalizability of the results. Additionally, two short-term experiments were conducted to simulate common office events that increase indoor air pollutant concentrations: human sneezing, and coughing. The impact of sensor positioning on more hazardous IAQ scenarios was evaluated across two desk locations with differing airflow conditions.

A Principal Component Analysis (PCA) was employed to identify the most suitable sensor package position. This position was determined based on the longitudinal data obtained from the five positions, three desks, and two orientations. The identified position was the one that contributed the most to the variance in the measurements, capturing the most information in relation to the occupant. Furthermore, Spearman's rank correlation coefficient ($\rho$), $R^2$, and Mean Absolute Error (MAE) were calculated for each position in comparison to the ground truth data collected near the particle dispersion machine's nozzle (the "breathing zone"). These calculations were performed to identify the position that exhibited the best correspondence in measurements during the short-term experiment. Key findings are presented as follows:

1) Sensor positions for personalized monitoring of different IEQ metrics vary within the indoor environment as different IEQ metrics are influenced by various factors and phenomena. For the specific conditions examined in this study, temperature, $CO_2$, and sound collected below the computer screen are more likely to capture conditions as they are experienced by the occupants in the workstation. Comparatively, humidity, $PM_1$, $PM_{2.5}$, $PM_{10}$ measurements were best captured above the monitor. Illuminance values were most accurately represented by a sensor located on the right side of the desk. Other configurations may yield different results in environments with varying spatial and environmental conditions.

2) Desk location, characterized by variations in lighting, thermal conditions, and airflow patterns in the monitored office, and the orientation of the sensors influence the selection of the optimal position. However, across all monitored desk locations, the sensor package facing the occupant consistently captured IEQ conditions as experienced by the occupant in the workstation.

3) Placing the sensor package above the monitor, at 0.7 m from the occupant and 1.2 m above ground, represents the best compromise for capturing temperature, humidity, $CO_2$, $PM_1$, $PM_{2.5}$, $PM_{10}$, illuminance, and sound under the conditions tested. This position also seamlessly integrates into the workspace without disrupting desk-related tasks. This not only enhances the visual appeal and ergonomics but also maintains a tidy desk, reducing obstructions in the occupant's workspace while ensuring the continued relevance of IEQ measurements.

While the findings provide actionable insights for effective IEQ sensor placement, it is important to note that results may vary under different office layouts and environmental conditions. In light of the absence of comprehensive guidelines, these insights can assist building practitioners in deploying IEQ sensing packages effectively. Building management and control systems can leverage reliable sensor data to adjust environmental conditions, maintain personalized IEQ levels, and improve occupant satisfaction.

## Acknowledgments

This material is based upon work supported by the Singapore Ministry of Education under grant no A-0008302-02-00 and A-8000136-01-00. Any opinions, findings, and conclusions expressed in this material are those of the authors and do not necessarily reflect the views of the Singapore Ministry of Education.